\documentclass[graybox]{svmult}

\usepackage{mathptmx}       
\usepackage{helvet}         
\usepackage{courier}        
\usepackage{type1cm}        
\usepackage{amssymb}
\usepackage{amsmath} 
\usepackage{makeidx}         
\usepackage{graphicx}        
\usepackage{multicol}        
\usepackage[bottom]{footmisc}

\makeindex             
                       
\makeatletter
\renewcommand{\fnum@figure}{\textbf{\figurename~\thefigure}}
\renewcommand{\fnum@table}{\textbf{\tablename~\thetable}}

\newcommand{\ba}{\begin{array}{llll}}
\newcommand{\ea}{\end{array}}

\newcommand{\be}{\begin{equation}\textstyle}
\newcommand{\ee}{\end{equation}}

\newcommand{\bi}{\begin{itemize}}
\newcommand{\ei}{\end{itemize}}

\newcommand{\bc}{\begin{center}}
\newcommand{\ec}{\end{center}}

\newcommand{\bfig}{\begin{figure}[!ht]}
\newcommand{\efig}{\end{figure}}

\newcommand{\ben}{\begin{enumerate}}
\newcommand{\een}{\end{enumerate}}

\newcommand{\bmat}{\left[\begin{matrix}}
\newcommand{\emat}{\end{matrix}\right]}

\begin{document}

\title*{Jam avoidance with autonomous systems}
\author{Antoine Tordeux \and Sylvain Lassarre}
\institute{Antoine Tordeux \at J\"ulich Supercomputing Centre, Forschungszentrum J\"ulich, Germany and Computer Simulation for Fire Safety and Pedestrian Traffic, Bergische Universit\"at Wuppertal, Germany, \email{a.tordeux@fz-juelich.de}
\and Sylvain Lassarre \at GRETTIA/COSYS -- IFSTTAR, France \email{sylvain.lassarre@ifsttar.fr}}
%
%
\maketitle

\abstract*{}

\abstract{
Many car-following models are developed for jam avoidance in highways. 
Two mechanisms are used to improve the stability: 
feedback control with autonomous models and 
increasing of the interaction within cooperative ones.  
In this paper, we compare the linear autonomous and collective optimal velocity (OV) models.
We observe that the stability is significantly increased by adding predecessors in interaction with collective models. 
Yet autonomous and collective approaches are close when the speed difference term is taking into account.   
Within the linear OV models tested, the autonomous models including speed difference are sufficient to maximise the stability.}

\section{Introduction} \label{intro}
Recently, many car-following models have been developed for jam avoidance in highways. 
Models have equilibrium homogeneous solutions where all vehicle speeds and spacings are constant and equal. 
`Jam avoidance property' is investigated through analysis of the stability of such solutions. 
Most of the approaches are extended versions of the \textit{optimal velocity} (OV) model \cite{ovm}. 
The basic model is solely based on the distance spacing with the predecessor (local next-neighbour interaction). 
Several studies shown that speed and spacing feedback mechanisms in autonomous OV models allow to improve the stability of the homogeneous solution 
and to avoid jam formation \cite{Konishi1999,Konishi2000,Davis2004,Zhao2005,Jin2013}. 
Similar results are obtained with the \textit{intelligent driver} (ID) model for specific parameter values \cite{Kesting2007,Kesting2008}. 

Several vehicles in the neighbourhood are taken in the interaction for collective (or cooperative) systems. 
Many studies show improvements of the stability if the number of predecessors in interaction increases \cite{Lenz1999,Hasebe2003,Wilson2004,Hu2014}.  
Comparable results are obtained with symmetric interaction
(interaction with predecessors and followers, see for instance \cite{Nakayama2001,Monteil2014}). 
Oppositely to autonomous models for which the variables can be directly measured, cooperative systems require that the vehicles are connected to communicate their states. 
This makes difficult their implementation. 

In this paper, autonomous linear OV models and extended ones with speed difference term are compared to their collective versions 
including several predecessors in interaction. 
Both extended and collective OV models describe significant improvement of the stability. 
More precisely, we observe that the number of predecessors in interaction in the collective models and 
the speed difference term in the autonomous approaches have similar roles in the dynamics. 
The paper is organized as follow. 
The linear jam avoidance models are introduced in section~\ref{def}.
The results of simulation experiment of a jam are presented in section~\ref{sim}, 
while the Lyapunov exponents of the different autonomous and collective models are calculated in section~\ref{exp}. 
The section~\ref{ccl} gives conclusion and outlook.

\section{Linear jam avoidance models}\label{def}

The optimal velocity model is
\be
\ddot x_n(t)=\frac1T\big(V(\Delta x_n(t))-\dot x_n(t)\big),
\label{ovm}
\ee
with $x_n(t)$ the position of the vehicle $n$ at time $t$, 
$\Delta x(t)=x_{n+1}(t)-x_n(t)$ the distance spacing with 
$x_{n+1}>x_n$ the predecessor position (see figure~\ref{scheme}), and $T>0$ the relaxation (or reaction) time  \cite{ovm}. 
A jam avoidance should have stable homogeneous solution. 
More precisely it should be locally stable with no oscillation (LSNO) to avoid collision and globally stable (GS), 
see for instance \cite[Chapter 15]{bookTreiber}. 
The conditions ensure a collision-free convergence of the system to the homogeneous solution for any initial condition. 
With OV model the OV model, the linear LSNO and GS conditions are respectively~:
\be
V'<\frac1{4T}\qquad\mbox{and}\qquad V'<\frac1{2T}.
\ee
Note that the first condition implies the second. 
The \textit{full velocity difference} (FVD) is an extended OV model including speed difference term \cite{Jiang2001}~:
\be
\ddot x_n(t)=\frac1{T_1}\big(V(\Delta x_n(t))-\dot x_n(t)\big)+\frac1{T_2}\Delta \dot x_{n}(t).
\label{fvdm}
\ee
It includes two relaxation times $T_1>0$ and $T_2>0$. 
The model is the same as the OV model at the limit $T_2\rightarrow\infty$. 
For the FVD model, the LSNO and GS conditions are respectively~:
\be
V'<\frac1{4T_1}\left(1+\frac{T_1}{T_2}\right)^2 \qquad\mbox{and}\qquad V'<\frac1{2T_1}+\frac1{T_2}.
\ee
These conditions are simply $V'<1/T$ if $T_1=T_2=T$ (the first inequality implies the second if $T_1<3T_2$). 
Clearly, the speed difference has a stabilization effect on the dynamics. 
The LSNO and GS conditions always hold at the limit $T_2\rightarrow 0$.

\bfig\bc%
\vspace{-25mm}\includegraphics{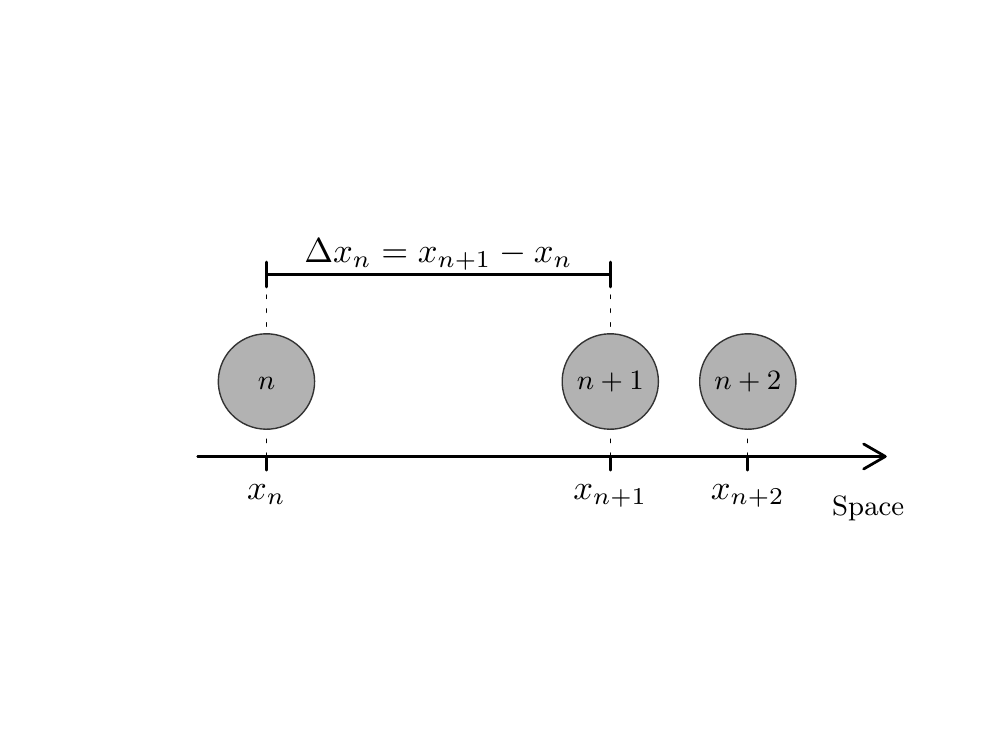}\vspace{-22.5mm}
\caption{Notations used. $x_n$ is the position and $\Delta x_n$ is the spacing of the vehicle $n$.}
\label{scheme}
\ec\efig

The models (\ref{ovm}) and (\ref{fvdm}) are autonomous ones~: 
they are solely based on distance spacing and speed difference with the predecessor. 
Collective models depend on several predecessors in front. 
Generally,  collective OV models have the form $\ddot x_n(t)=\sum_{k=1}^K F_k\big(\Delta x_{n,k}(t),\dot x_n(t),\Delta \dot x_{n+k}(t)$,
where $K$ is the number of predecessors taking into account and $\Delta x_{n,k}=x_{n+k}-x_k$ is the distance to the vehicle $n+k$. 
$F_k$ represents the influence of the vehicle $k$ on the acceleration rate of the considered vehicle. 
In the \textit{Multi-anticipative} (MA) model \cite{Lenz1999}, this influence is
\be 
F_k =\frac{\alpha _k}T\big(V\left(\Delta x_{n,k}(t)/k\right)-\dot x_n(t)\big).
\label{mam}
\ee
The \textit{velocity difference multi-anticipative} (VDMA) model includes speed difference terms
\be
F_k =\alpha _k \left[\frac1{T_1}\big(V\left(\Delta x_{n,k}(t)/k\right)-\dot x_n\big) + 
\frac1{T_2}\Delta\dot x_{n+k}(t)\right].
\label{vdmam}
\ee
Here the positive coefficients $(\alpha_k)$ are such that $\sum_k\alpha_k=1$. 
They specify the interaction with the predecessors. 
In the following we set $\alpha_k=1/K$ for all $k$ (uniform interaction) in order to maximise the stability  \cite{Lenz1999,Hasebe2003}. 
Note that the MA model is the OV one and the VDMA model is the FVD one for $K=1$, while the VDMA model is the MA one at the limit 
$T_2\rightarrow\infty$. 
The tested models are resumed is table~\ref{res}. 
\begin{table}[!ht]\bc\begin{tabular}{l|l|l|l}
Name&$\ $Acronym$\ $&$\ $Type&$\ $Parameter\\[1mm]
\hline
&&&\\[-2mm]
Optimal velocity&$\ $OV&$\ $Autonomous$\ $&$\ V'$, $T$\\[0mm]
Full velocity difference&$\ $FVD$\ $&$\ $Autonomous&$\ V'$, $T_1$, $T_2$\\[0mm]
Multi-anticipative&$\ $MA&$\ $Collective&$\ V'$, $T$, $K$\\[0mm]
Velocity difference multi-anticipative$\ $&$\ $VDMA&$\ $Collective&$\ V'$, $T_1$, $T_2$, $K$
\end{tabular} \caption{Name, acronym, type and parameters of the tested models.}\label{res}\vspace{-2mm}\ec\end{table}

\section{Simulation of a jam} \label{sim}
\begin{figure}[!ht]\bc\vspace{-7mm}
\includegraphics[width=6.5cm]{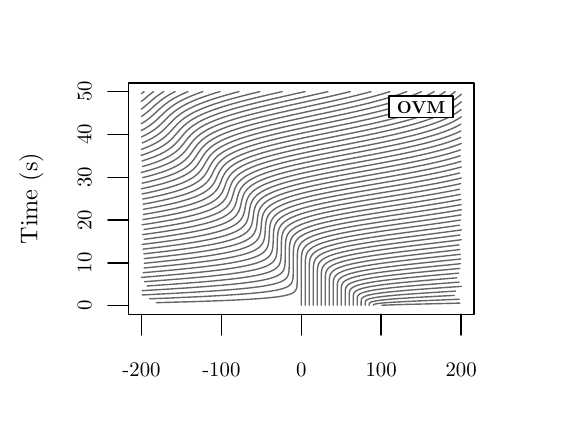}\hspace{-15mm}\includegraphics[width=6.5cm]{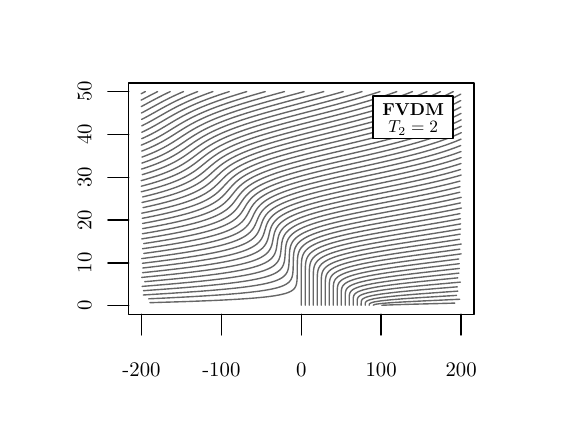}\\[-15mm]
\includegraphics[width=6.5cm]{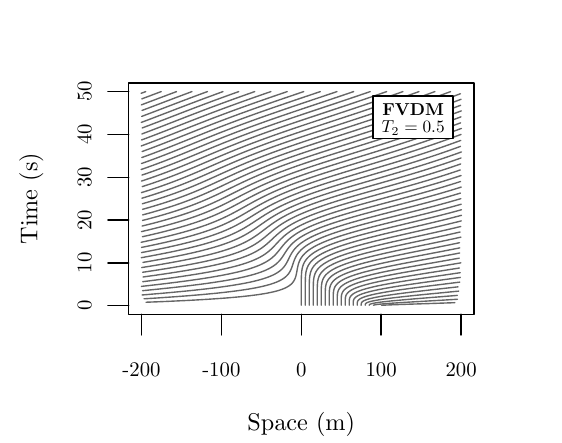}\hspace{-15mm}\includegraphics[width=6.5cm]{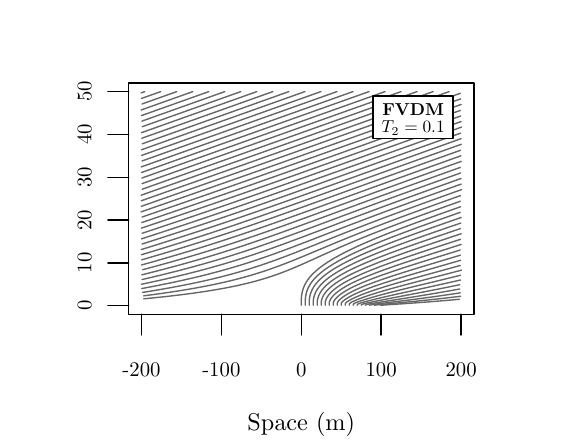}
\caption{Trajectories with the OV and FVD models from jam initial configuration.}
\label{trajFVDM}
\ec\end{figure}

\begin{figure}[!ht]\bc\vspace{-5mm}
\includegraphics[width=6.5cm]{trajOVM.pdf}\hspace{-15mm}\includegraphics[width=6.5cm]{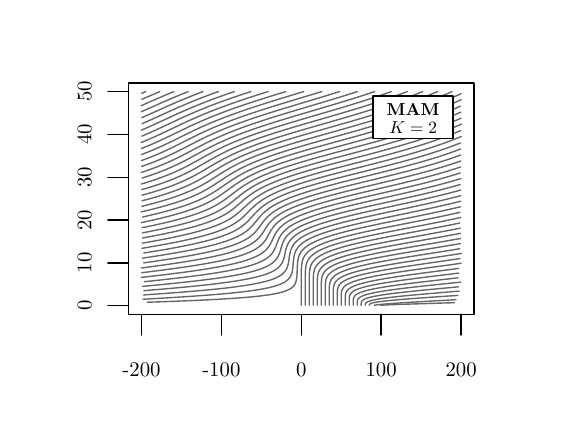}\\[-15mm]
\includegraphics[width=6.5cm]{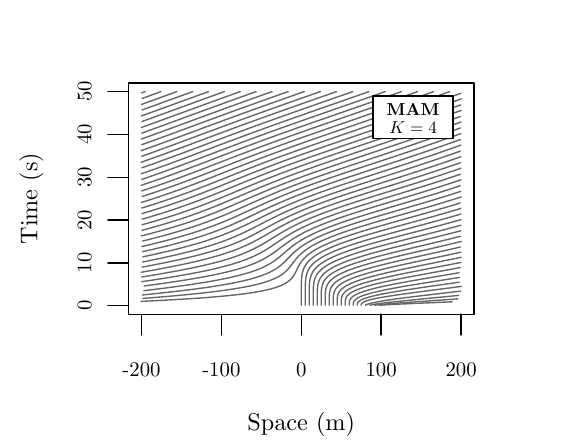}\hspace{-15mm}\includegraphics[width=6.5cm]{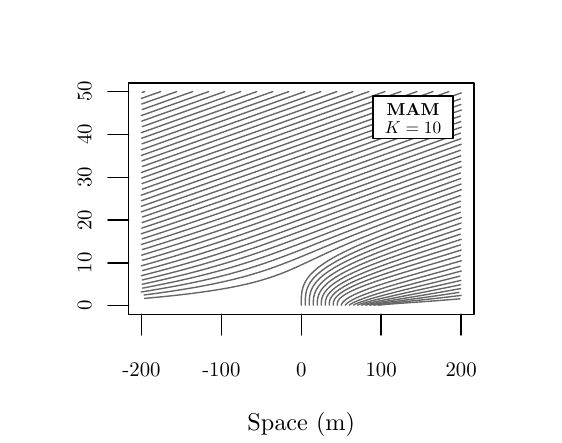}
\caption{Trajectories with the OV and MA models from jam initial configuration.}
\label{trajMAM}\vspace{0mm}
\ec\end{figure}

\begin{figure}[!ht]\bc\vspace{-5mm}
\includegraphics[width=6.5cm]{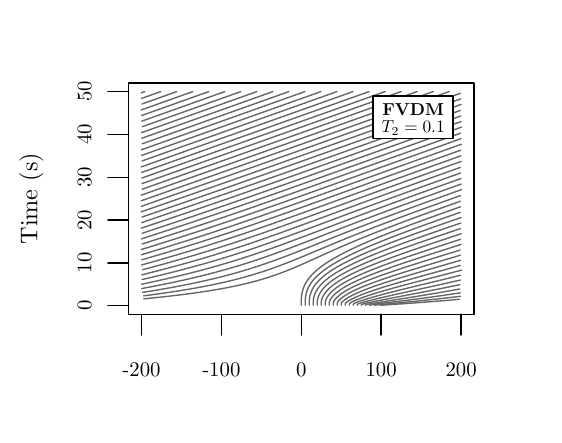}\hspace{-15mm}\includegraphics[width=6.5cm]{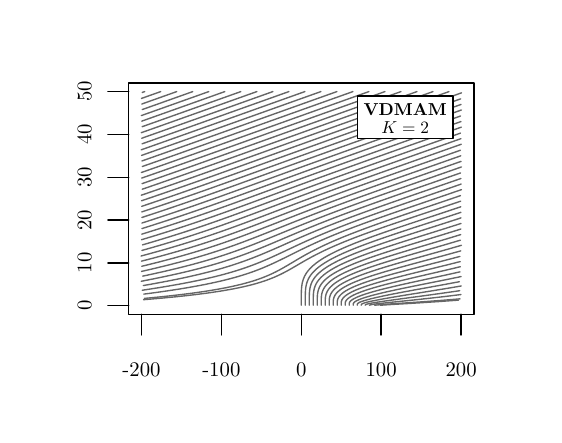}\\[-15mm]
\includegraphics[width=6.5cm]{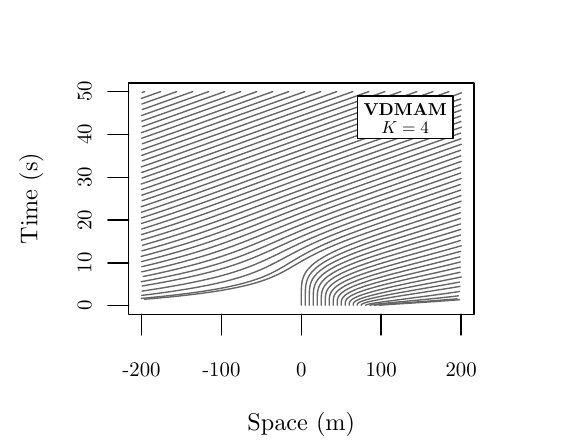}\hspace{-15mm}\includegraphics[width=6.5cm]{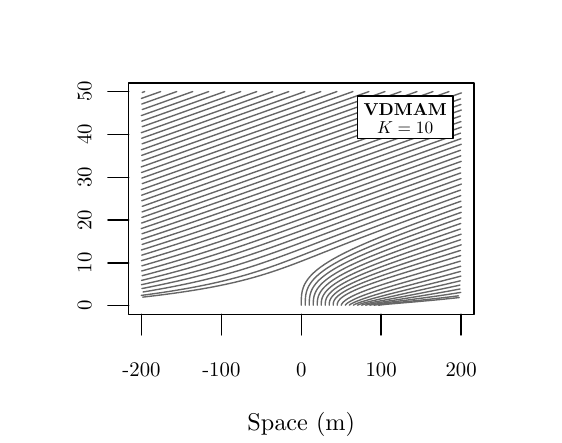}
\caption{Trajectories with the FVD and VDMA models from jam initial configuration.}
\label{trajVDMAM}\vspace{-10mm}
\ec\end{figure}

In this section, the models (\ref{ovm}), (\ref{fvdm}), (\ref{mam}) and (\ref{vdmam}) are simulated 
with periodic boundary conditions from jam initial conditions by using explicit Euler schemes with time step 0.001 s. 
$N=20$ vehicles are considered with the settings: $V'=1\ $s$^{-1}$~, $T=T_1=0.25\ $s (fix), and $T_2=2$, $0.5$, $0.1\ $s~, $K=2$, $4$, $10\ $veh (tested). 
The settings are defined so that the LSNO and GS conditions occur for any model. 
The trajectories obtained with OV and FVD autonomous models are presented in figure~\ref{trajFVDM}. 
The convergence speed to the homogeneous solutions increases as $T_2\rightarrow0$. 
The same phenomenon occurs with MA model as $K\rightarrow\infty$, see figure~\ref{trajMAM}. 
However, there is no clear improvements of the stability with the VDMA model 
if $T_2$ is sufficiently small (see figure~\ref{trajVDMAM}).

The speed of convergence of the system to the uniform solution can be quantified by spacing standard deviation sequence (Lyapunov function)~: 
\be\textstyle
\sigma_{\Delta x}=\sqrt{\frac1N\sum_{n=1}^N\left(\Delta x_n-\Delta \bar{x}_n\right)^2}\qquad\mbox{with}\quad \Delta \bar{x}_n=\frac1N\sum_{n=1}^N\Delta x_n.
\ee 
In the figure~\ref{stddev}, the logarithms of the spacing standard deviation are plotted according to the time for the different models. 
We observe linear evolution, meaning that the deviation tend to zero with exponential speed. 
As expected, the slope of the logarithm (i.e. the convergence speed) increases as $T_2$ decreases with the autonomous models 
(see figure~\ref{stddev}, top left panel), while the speed depends on the number of predecessors in interaction $K$ 
with the collective MA model (see figure~\ref{stddev}, top right panel). 
As we observed previously, the speed does not change significantly if $K$ increases with VDMA model (see figure~\ref{stddev}, bottom left panel). 
In fact the speeds of convergence of FVD, MA, and VDMA models are close (see figure~\ref{stddev}, bottom right panel); 
they are strongly faster than the convergence speed of ordinary OV model. 
Such results suggest that speed difference term with the autonomous models and the number of predecessors in interaction with the collective ones  
have similar roles in the dynamics. 
The convergence speed to the homogeneous solution is maximised as $T_2\rightarrow0$ or as $K\rightarrow\infty$. 

\bfig\bc\vspace{-7mm}
\includegraphics[width=6cm]{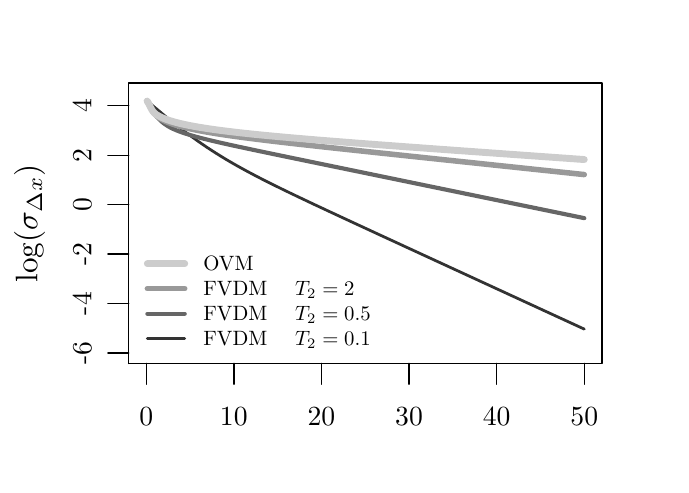}\hspace{-10mm}\includegraphics[width=6cm]{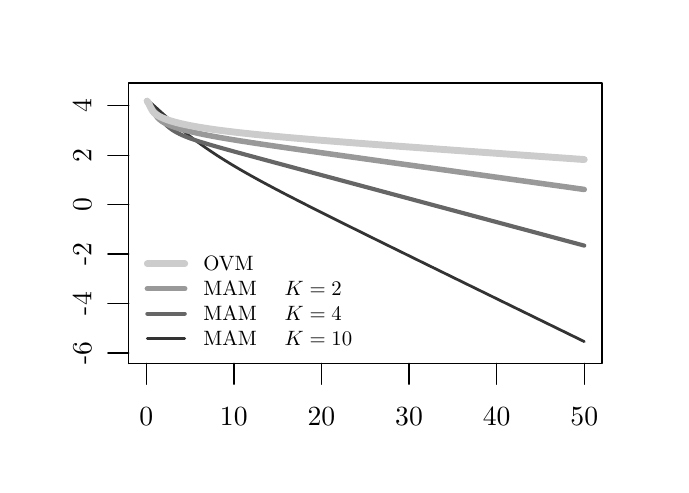}\\[-10mm]
\includegraphics[width=6cm]{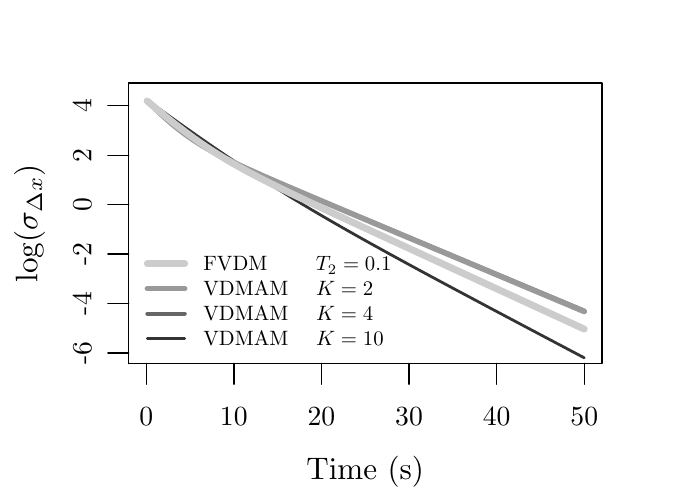}\hspace{-10mm}\includegraphics[width=6cm]{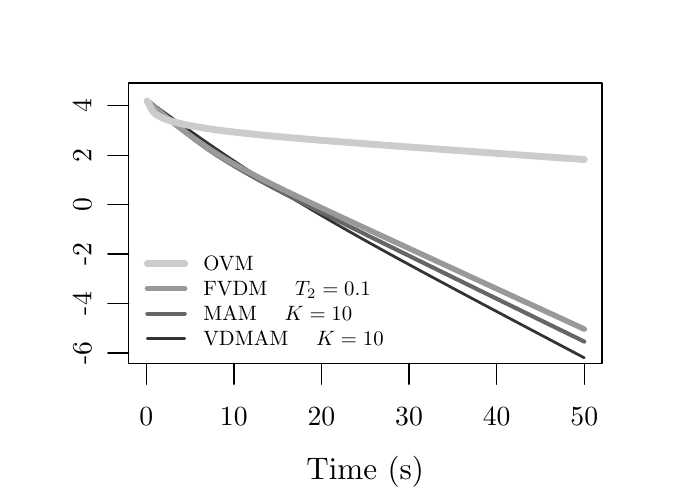}
\caption{Sequences of the spacing standard deviation logarithm with OV, FVD, MA and VDMA models.}
\label{stddev}\vspace{-12mm}
\ec\efig

\section{Lyapunov exponents} \label{exp}

The solution of the linear systems are linear combination (LC) of exponential terms
\be
x_n(t)=\mbox{LC}\big(\exp(\lambda_lt),t\exp(\lambda_lt)\big)
\ee
with $(\lambda_l)$ the Lyapunov exponents of the system (i.e. the eigenvalues of the system Jacobian matrix). 
In our stable case, all the exponents have strictly negative real parts, excepted one equal to zero. 
Moreover, we can expect than the convergence to the homogeneous solution gets faster as the exponents go the 
left of the imaginary axis. 
With the optimal velocity we investigate, the exponents are~:  
\be
\lambda_l=\frac12\sum_{k=0}^K\beta_k \iota_l^k\pm\frac12\Big[\left(\sum_{k=0}^K\beta_k\iota_l^k\right)^2-4\sum_{k=1}^K\alpha_k(1-\iota_l^k)\Big]^{1/2}
\ee
with $\iota_l=\exp(2i\pi l/N)$, $l=1,\ldots N$, $N$ being the vehicles number, 
$\alpha_k=\frac{1}{kT_1}\frac{V'}{K}$, $\beta_0=-\frac1{T_1}-\frac1{T_2}$ and 
$\beta_k=-\frac1{KT_2}$ for all $k=1,\ldots N$. 
The Lyapunov exponents are plotted in figure~\ref{exp1} for the different models. 
We observe that they converge to a double mode pattern as $T_2\rightarrow0$ with the autonomous FVD model, 
and $K\rightarrow\infty$ with the MA collective model. 
They remain double mode with the collective VDMA model as $K$ increases. 
Such results confirm qualitatively the ones observed by simulation.
The speed difference behave in the dynamics as the number of predecessors in interaction.
Also increasing the interaction seems not necessary to maximise the stability. 


\bfig\bc\vspace{-0mm}
\includegraphics[width=\textwidth]{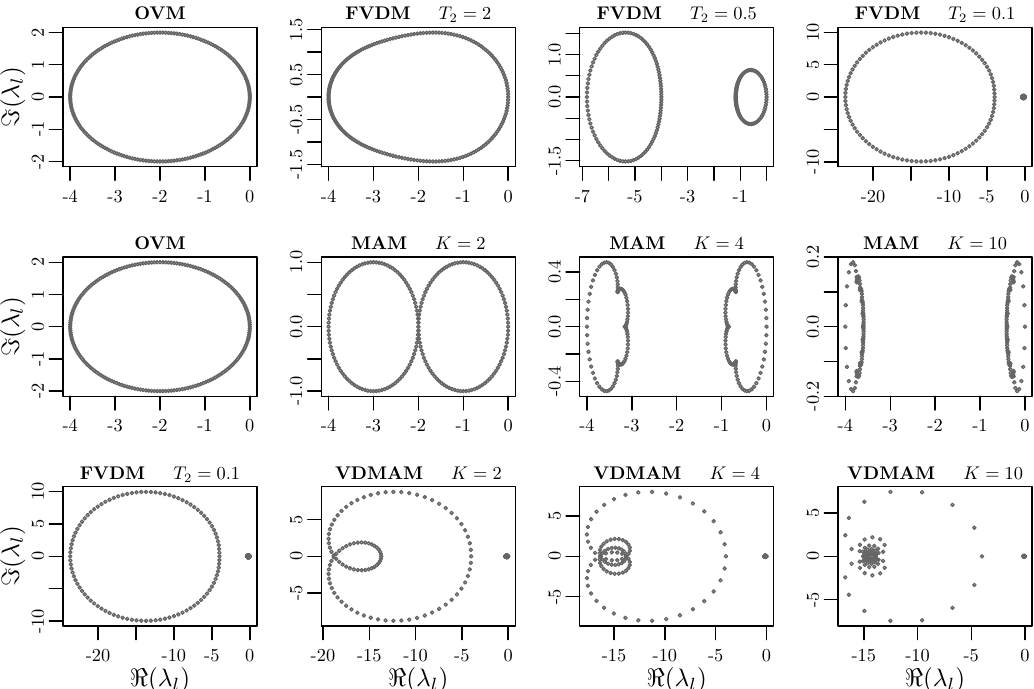}
\caption{Lyapunov exponents for OV, FVD (top panels), MA  (middle panels) and VDMA (bottom panels) models with $N=100$.}
\label{exp1}\vspace{-12mm}
\ec\efig


\section{Conclusion} \label{ccl}

The convergences to the homogeneous solution of linear jam avoidance OV models are compared. 
We observed that extending the OV model with speed difference term
significantly improves the stability. 
In a similar way, the adding of neighbours in interaction gives stability enhancements. 
However, increasing the interaction does not improve the stability with the extended OV model. 
This suggests that both the number of predecessors in interaction in the collective models and 
the speed difference term in the autonomous approaches allow to maximise the convergence speed to homogeneous solutions. 
Also, the connection between the vehicles, hard to implement, may not be necessary to optimise the stability 
and efficiently avoid jam formation.
Further investigations remain to be carried out to validate this hypothesis. 
For instance, the influence of the geometry, initial conditions or vehicle density have to be investigated.
The shape of the Lyapunov exponents and their impact on the stability are not explicit. 
Furthermore, non-linear models may present better convergence speed than the basic linear models we analysed. 
These subjects will the topic of future works.

\bibliographystyle{spmpsci}
\bibliography{bibli2}

\end{document}